%% file: RMHpreprint.tex
\documentclass[12pt,preprint]{aastex}

\usepackage{graphicx} 
\usepackage{longtable,lscape}

\slugcomment{To appear in the Astrophysical Journal}
\shorttitle{M31 and M33 Hypergiants}
\shortauthors{Humphreys et al. }

\begin{document}

\title{Luminous and Variable Stars in M31 and M33. I. The Warm Hypergiants and
Post-Red Supergiant Evolution\altaffilmark{1}}

\author{
Roberta M. Humphreys,\altaffilmark{2}
Kris Davidson,\altaffilmark{2} 
Skyler Grammer,\altaffilmark{2}
Nathan Kneeland,\altaffilmark{2} 
John C. Martin,\altaffilmark{3}  
Kerstin Weis,\altaffilmark{4} 
and
Birgitta Burggraf,\altaffilmark{4}
}

\altaffiltext{1}  
{Based  on observations  with the Multiple Mirror Telescope, a joint facility of the Smithsonian Institution and the University of Arizona and on observations obtained with the Large Binocular Telescope (LBT), an international collaboration among institutions in the United
States, Italy and Germany. LBT Corporation partners are: The University of
Arizona on behalf of the Arizona university system; Istituto Nazionale di
Astrofisica, Italy; LBT Beteiligungsgesellschaft, Germany, representing the
Max-Planck Society, the Astrophysical Institute Potsdam, and Heidelberg
University; The Ohio State University, and The Research Corporation, on
behalf of The University of Notre Dame, University of Minnesota and
University of Virginia.
} 

\altaffiltext{2}
{Minnesota Institute for Astrophysics, 116 Church St SE, University of Minnesota
, Minneapolis, MN 55455; roberta@umn.edu} 

\altaffiltext{3}
{University of Illinois, Springfield, IL}
\altaffiltext{4}
{Astronomical Institute, Ruhr-Universitaet Bochum, Germany,}

\begin{abstract}
The progenitors of Type IIP supernovae have an apparent  upper limit to their 
initial masses 
of about 20 M$_{\odot}$, 
suggesting that the most massive red supergiants evolve to warmer temperatures before their 
terminal explosion. But very few post-red supergiants are known. We have identified a small group of luminous stars in M31 and M33 that are candidates for post-red 
supergiant evolution. These stars have A -- F-type supergiant absorption line spectra and strong 
hydrogen emission. Their spectra are also distinguished by the 
Ca II triplet and [Ca II] doublet in emission formed in a low density circumstellar environment. 
They all have significant near- and mid-infrared
excess radiation due to free-free emission and thermal emission from dust. We estimate the 
amount of mass 
they have shed and discuss their 
wind parameters and mass loss rates which range from a few $\times$ 10$^{-6}$  to 
10$^{-4}$ M$_{\odot}$ yr$^{-1}$. On an HR Diagram, these stars will overlap the region of the 
LBVs at maximum light, however the warm hypergiants are not LBVs. Their non-spherical 
winds are not optically 
thick and they have not exhibited any  significant variability. We suggest, however, that the warm hypergiants may be the progenitors of the ``less luminous'' LBVs such as R71 and even SN1987A. 
\end{abstract} 


\section{Introduction}
The progenitors of Type IIP supernovae have an apparent  initial mass upper limit of 
about  20 solar masses \citep{Smart},  suggesting that red supergiants with higher 
initial masses  do not explode as Type IIP SNe. Thus the most luminous
red supergiants may evolve to warmer temperatures before their terminal explosions.  
 This leaves us with the problem of where are these post-red supergiants and how do we recognize them?
 This is an important question for understanding the final stages of massive star evolution.

As part of a larger program on the luminous and variable stars in M31 and M33, we have identified
a small subset of stars which we call warm hypergiants which  may be post-red supergiants. 
In this first paper on the M31 and M33 stars, we describe  
 the warm hypergiants and the evidence that they may be examples of post-red supergiants. 
 In Paper II we will present the results of our spectroscopic 
survey of LBVs, candidate LBVs, emission line stars and other supergiants 
in M31 and M33, and in Paper III we will discuss the evidence for variability
and instability in these stars. 

In the next section we describe our new observations. Inspection of the spectra
quickly revealed a  few stars in each galaxy with A to F-type supergiant 
spectra and strong hydrogen emission. Their distinguishing characteristics, however,  
are  the presence of the Ca II triplet near 8500{\AA} and [Ca II] doublet near 7300{\AA} 
in emission, indicating  
 extensive circumstellar ejecta plus excess infrared radiation.  
The  individual stars are described in \S {3}, and 
in \S {4} we present the evidence  
for  circumstellar nebulae, dusty ejecta, and significant mass 
loss in these stars. Their evolutionary state  is discussed in the last section.

\section{Target Selection and Spectroscopy}

We selected known LBVs and LBV candidates, luminous emission-line stars, and 
evolved variables  
from the survey by \citet{Massey07}, an unpublished H$\alpha$ survey by Weis and collaborators, and the list of candidates from \citet{Valeev10a}. The observations  were 
made in October 2010 with the Hectospec Multi-Object Spectrograph (MOS) \citep{Fab98} on the 
6.5-m MMT on Mt. Hopkins. The Hectospec\footnote{http://www.cfa.harvard.edu/mmti/hectospec.html} has a 1$\arcdeg$ FOV and uses 300 fibers each with
a core diameter of 250$\mu$m subtending 1$\farcs$5 on the sky. Fibers were assigned to 27 stars in M31 and 57 in M33. We used the 600 l/mm 
grating with the 4800{\AA} tilt  yielding  $\approx$ 2500{\AA} coverage with 0.54{\AA}/pixel resolution and R of $\sim$  2000. The same grating with a tilt of 6800{\AA} was used for the red specta with $\approx$ 2500{\AA} coverage, 0.54{\AA}/pixel resolution and R of 
$\sim$ 3600.  The blue and red spectra were  
reduced using an exportable version of the CfA/SAO SPECROAD package 
for  Hectospec data\footnote{External SPECROAD was developed by 
Juan Cabanela for use
on Linux or MacOS X systems outside of CfA. It is available online at:
http://iparrizar.mnstate.edu.}. The spectra were all bias subtracted, flatfielded and 
wavelength calibrated. Because of crowding, the sky subtraction was done using 
fibers assigned outside the field of the galaxy. For this reason, some of the spectra exhibit obvious nebular contamination. A journal of the observations  is included in Table 1. 

A few stars of special interest (5 in M31 and 8 in M33) were also observed with the MODS1 spectrograph 
on the Large Binocular Telescope (LBT) during commissioning in September 2011, 
and in October and November 2012, and January 2013. The MODS1 uses a dichroic to obtain 
blue and red spectra simultaneously with the  G400L and G750L gratings   
for the blue and red channels, respectively, yielding  wavelength coverage from  3200{\AA} in the blue to more than 1$\mu$m in the
red. With a 0\farcs6 slit the resolution is   $\sim$ 2000. 
The two dimensional images  were initially reduced by R. W. Pogge in 2011 
following standard procedures for bias subtraction and flat fielding. The spectra from 2012 and 2013 were similarly reduced using software provided by R. W. Pogge for 
the MODS spectra The spectra were then extracted and  
wavelength, and flux calibrated using the IRAF twodspec
and onedspec packages. The observations of these individual stars are also listed in 
Table 1.

\section{The Warm Hypergiants}

The warm hypergiants are  an especially interesting and significant class of objects for understanding
the final stages and pre-SN evolution of massive stars. These stars, like IRC+10420  \citep{Jones,Rene96,Rene98,RMH97,RMH02}, $\rho$ Cas \citep{deJ98,Lobel,Gorlova}, and HR8752 \citep{Hans} in the Milky Way,  are most likely on a post-red supergiant blue loop to warmer 
temperatures.  The warm hypergiants in M31 and M33  include the peculiar Var A in M33 \citep{RMH87,RMH06} and B324 in M33 
\citep{RMH80,HMF90,Massey96,Clark12}, one of its visually brightest stars.

Based on their absorption lines, these stars  have A to F-type
supergiant spectra and strong hydrogen emission, hence the ``warm'' hypergiant name. 
The A -- F spectrum may correspond to the star's actual photosphere or it may  
originate in an  
optically thick wind, see \S {4}.  As discussed below several of these stars show spectroscopic evidence for  winds and mass loss. In either case, the absorption-line spectrum represents  the region from which the visual  light is escaping. A critical characteristic of these stars is the presence of the near-infrared Ca II triplet in emission and the rare [Ca II]
$\lambda$7291 and $\lambda$7323 emission lines. The [\ion{Ca}{2}]  doublet arises
from the lower level of the transition that produces the triplet ($\lambda\lambda$8498,8542,8662) emission lines, which are formed
in the star's ejecta   by radiative de-excitation from the strong  \ion{Ca}{2} H
and K absorption upper levels. The [\ion{Ca}{2}]
levels are  normally collisionally de-excited back to the ground state 
unless the density is sufficiently low.
The [\ion{Ca}{2}] emission lines thus imply 
a  low density  circumstellar medium. They are not commonly seen in
the cool, dense  winds of classical LBVs at maximum, possibly because
 \ion{Ca}{2} has a low ionization potential and is suppressed in the
presence of UV radiation which may be present in the environment from the hot
underlying star.  R71 in the LMC  did develop [Ca II] emission during 
its current maximum \citep{Gamen,Mehner}, the first recorded for an LBV in eruption. 
R71, however,  may be an example of a giant eruption LBV \citep{Mehner}. 

\citet{Massey07} called some of these warm hypergiants  ``hot LBV candidates'' and in one
case a ``P Cyg LBV candidate''. The presence of Ca II H and K, other 
absorption lines common in A to F-type supergiants, the luminosity sensitive O I triplet at $\lambda$7774{\AA} 
in A and F-type supergiants,  and the Ca II and [\ion{Ca}{2}] emission
lines, however, indicate that they  are  evolved, cooler stars. 
The individual stars are discussed below with examples of their
blue and red spectra.

Although {\it Variable A} in M33 was one of the original Hubble-Sandage variables \citep{HS}, it is not an LBV. It is one of the very luminous, unstable stars that define
the empirical upper luminosity boundary in the HR Diagram, see \citet{HD94}. It  experienced a
high mass loss episode that lasted about 45 years.
Its photometric and spectroscopic
variablility has been described in some detail by Humphreys et al. (1987, 2006).
Briefly, in 1951, when it was
one of the visually brightest stars in M33,  it rapidly
{\it declined}  by 3  mag.  We now know that this was due to a shift
in its energy distribution to lower temperatures and to the
formation of dust. In 1985-86, Var A had the spectrum of an M supergiant and  a
large infrared excess at 10$\mu$m. Its large spectral and photometric variations had occurred at nearly constant bolometric luminosity. Twenty years later, its spectrum had returned to a warmer F-type photosphere consistent with its properties at maximum light 50 years earlier \citep{RMH06}. Although its colors have likewise shifted
bluer, Var A has remained faint, but see Table 3 for an update on the magnitudes.  Comparison with our spectra from 2003-04 shows
little change, though the hydrogen emission may have weakened slightly. Our current spectrum from September  2011 shows that it still has a late
F-type spectrum and  prominent emission lines of Ca II in the near-infrared,
[Ca II], and relatively strong lines of K I. The importance of these lines was 
 discussed in \citet{RMH06}.  The higher
signal-to-noise Hectospec and MODS1 spectra also  permit a more accurate estimate of its apparent
spectral type as F8.

{\it B324} in M33 has been classified as A5 Iae \citep{RMH80,HMF90} and F0-F5 Ia
\citep{Mont}.  
Absorption lines in its MODS1 blue and red spectra closely resemble the current
spectrum of IRC~+10420 \citep{Rene98,RMH02}.
The Sr II $\lambda$4078 and $\lambda$4215 lines, and the strong Fe II, Ti II blends at $\lambda\lambda$4172,4178 indicate  very high luminosity. The
relative strengths of the
Ca II $\lambda$4227 and Fe II $\lambda$4233 lines and the N I absorption lines
in the red indicate that B324 is later than an early A-type
supergiant but not as cool  as a late F-type. Based on these lines we recommend a
spectral type of A8 - F0. H$\beta$ is in emission but the higher Balmer
lines are in absorption. Weak Fe II and Ti II emission  lines  also observed in IRC~+10420 are present at $\lambda$ $>$ 5000{\AA}.  The red spectrum shows strong H$\alpha$ emission with broad wings,  a strong O I triplet in absorption, plus the Ca II triplet and [Ca II] doublet in emission
indicating the presence of low density circumstellar material. The Ca II triplet
emission lines also  have P Cygni absorption components. 
 Several  Fe II emission lines plus strong Si II absorption at
$\lambda$6347{\AA} and $\lambda$6371{\AA} are also present between 6000{\AA}
and H$\alpha$.

\citet{Massey96} and more recently \citet{Clark12} have suggested on the
basis of possible spectral variability that B324 may be an LBV candidate or an LBV
in an extended
maximum light, cool dense wind state, i.e. an LBV eruption. However, that
conclusion is based on a limited apparent spectral type change (see above).
Its current spectrum from a high quality digital
spectrum is consistent with a very high luminosity late A-type supergiant.
Furthermore the strong Ca II and
[Ca II] emission and lack of variability cast  doubt on whether it is an LBV. Instead B324 shares
the spectral characteristics of IRC~+10420 and  other warm hypergiants.
B324 also has a small near-infrared excess  due to free-free emision in its wind (Figure 14). 
Clark et al. also argued that B324 was above the empirical upper
lumninosity boundary even though it was one of the stars that defined it
\citep{RMH83}.
Their conclusion however depended on a larger adopted distance modulus for M33
derived from an eclipsing binary \citep{Bonanos} which was 0.6 mag 
higher than that used by \citet{RMH83}. In this work we use the Cepheid distance scale, see \S {4.3}. 
The  luminosities of B324 and the other warm hypergiants are discussed
there.  

The MODS1 blue and red absorption line spectra of {\it N093351} and {\it N0125093}  resemble
those of B324 and IRC+10420.
Both spectra show the critical Ca II triplet and [Ca II] emissions.  The [Ca II]
lines in N093351 however are very strong relative to the Ca II triplet which have asymmetric
double emission  profiles with a weaker blue component.
 The H$\alpha$ emission line is also double with a weaker blue component.
Although the depths of the absorption minima in the Ca II lines and H$\alpha$ are
not as deep as in IRC~+10420, this is another characteristic it
shares with IRC~+10420 possibly  due to a bipolar outflow. The velocities of the blue and red emission components
relative to the absorption minima in the Ca II triplet lines indicate an
outflow of about $\pm$ 70 km s$^{-1}$ and in H$\alpha$ the relative velocities are  
$\pm$  120 km s$^{-1}$. The [Ca II] emission lines are single and have velocities
that agree with the absorption minima in the Ca II triplet profiles.  The H$\alpha$ 
profile in N125093 also has an additional emission component on its blue side
corresponding to an outflow velocity of $\approx$ $\pm$ 200 km s$^{-1}$. The H$\alpha$ and H$\beta$ line profiles 
in both stars have broad wings with  a P Cygni absorption feature at H$\beta$. 
Both stars 
also have an  infrared excess due to circumstellar dust. 
 N093351 and  N0125093 have been  proposed to be LBVs \citep{Valeev09,Valeev10b},
 but their spectral characteristics, infrared excess, and limited variability  
are more like IRC~+10420 and B324.

The blue and red MODS1 spectra of B324, N093351, N0125093, and Var A are shown in Figure 1. 
N093351's  double  H$\alpha$ and Ca II triplet profiles and the [Ca II] 
emission profiles are also shown separately in Figure 2.  
The H$\alpha$ and H$\beta$ profiles for  the warm hypergiants are discussed in the next section on Thomson scattering (\S {4.2}) and are shown in Figures 5 -- 10 in the on-line edition.

We have also identified three warm hypergiants in M31 with the  Ca II and [Ca II]
emission.  \citet{Massey07} called {\it M31-004444.52} a
P Cyg LBV candidate\footnote{In this paper we use the galaxy name plus the RA of the star as its designator.}. The Balmer and Fe II  emission lines do indeed show
deep P Cygni absorption features,  but it  is spectroscopically much more like IRC~+10420 than P Cyg.
 M31-004444.52 not only has the signature Ca II and [Ca II] emission,   but also strong O I $\lambda$7774 absorption, and Ca II H and K and 
other absorption lines characteristic of luminous A and F-type supergiants 
such as the Fe II, Ti II blends at 4172{\AA} and 4178{\AA}. Based on its 
absorption lines its apparent spectral type is $\approx$ F0. The hydrogen emission profiles are asymmetric and 
concave to the red with very broad wings, 
 the characteristic signature of Thomson scattering very likely in the star's wind (\S {4.2} and Figure 4), 
plus deep P Cygni absorption. The strong Fe II emission lines from multiplet 42 ($\lambda\lambda$4924,5018,5169) display the same asymmetric 
profiles from Thomson scattering with  P Cyg features. The Ca II and [Ca II] emission profiles are similarly asymmetric and the Ca II triplet lines also show P Cyg absorption.  

The spectrum of {\it M31-004522.58} shows the higher Balmer lines in absorption, the Ca II K line, He I 4026{\AA} and 4009{\AA} in absorption together with 
weak Fe II emission.  
These He I lines plus the relative strengths of the  He I $\lambda$4471{\AA} and Mg II $\lambda$4481{\AA} absorption 
lines suggest a somewhat earlier spectral type than for the other hypergiants, 
$\approx$ A2.  H$\alpha$ also has
a double emission line profile with broad wings resembling N093351 in M33. The relative
velocities of the blue and red emission components are $\approx$ $\pm$ 105 km s$^{-1}$ with respect to the absorption minimum. The Ca II triplet lines, however, are not double, although a higher resolution spectrum might show similarly double profiles. In addition to  Ca II and [CaII], O I 8446{\AA} is in emission. The O I blend at 7774{\AA} is very weakly 
present in absorption.  We would normally expect to see both O I lines in emission in about 
equal strength, but 8446{\AA} can be pumped by flourescence with Lyman $\beta$.  Thus O I  7774{\AA} is probably at least partially filled in by emission.  

Although the blue and red  Hectospec spectra  of {\it M31-004322.50} have 
low S/N, there are numerous 
absoption lines in the blue characteristic of late A or early F-type 
supergiants, a strong O I 7774{\AA} line in absorption and the [Ca II] doublet in emission. 
Its spectrum is very much like that of M31-004444.52. 
H$\alpha$ and H$\beta$ have asymmetric profiles with
broad red wings characteristic of Thomson scattering  and deep P Cygni absoption features. This asymmetry
is shared with the three strong Fe II emission lines  at 4924{\AA},
5018{\AA}, and 5169{\AA} from multiplet 42 also with P Cygni absorption. The Hectopsec red spectra do not go past 8000{\AA}, so we cannot confirm if the Ca II triplet is in emission.  
The blue and red spectra of the three warm hypergiants in M31 are shown in Figure 3,
and their H$\alpha$ and H$\beta$ emission profiles are shown in Figure 4 for M31-004444.52 and in Figures 5 -- 10, in the on-line edition.   

The warm hypergiants are listed in Table 2 with their positions and  apparent 
spectral types.  Only one of these stars, N125093 in M33, is the lists of candidate yellow 
supergiants by \citet{Drout,Drout2012}.  The visual photometry from \citet{Massey06} and near- and mid-infrared photomery from cross-identification with 2MASS \citep{Cutri},  the {\it Spitzer} surveys of M31 \citep{Mould} and M33 \citep{McQ,Thom} and WISE \citep{Wright} is summarized in Table 3. 
All of the warm hypergiants have excess infrared emission. Their resulting spectral energy distributions (SEDs), circumstellar ejecta, and mass loss indicators are discussed in the next section. 

\section{Circumstellar Nebulae, Winds, Dusty Ejecta,  and Mass Loss}

\subsection{The Gaseous Circumstellar Ejecta}

The spectra of the warm hypergiants display a variety of  indicators for  
extensive circumstellar gas, stellar winds, and mass loss. These  include  
deep P Cygni absorption profiles,    
 Thomson scattering wings formed in their stellar winds, and the Ca II, [Ca II] emission lines.
 Three of these 
stars  show double emission profiles in H$\alpha$, and in Ca II in one star, probably due to bipolar outflows.  The outflow velocities measured from the double 
profiles and the absorption minima in the P Cygni profiles are summarized in Table 4. 
The velocities   range from less than
100 km s$^{-1}$ up to $\sim$ 300 km s$^{-1}$ and are typical for A to F-type supergiants.  
We note that there is tendency for the velocities from the Ca II and [Ca II] emission lines to be somewhat lower 
than those for the hydrogen and Fe II lines in the same star. The Ca II and [Ca II] lines  
 may be formed in a different region than the other lines which can 
originate in denser gas closer to the star.  

We estimate the densities in the outflow from the ratio of the [Ca II] $\lambda$7300 and 
Ca II triplet circa $\lambda$ 8500 equivalent widths of the emission multiplets. The Ca II triplet emission is 
produced by radiative de-excitation from the absorption from the strong Ca II H and K lines ($4s \; ^2\mathrm{S}$ $\rightarrow$  $4p \; ^2\mathrm{P}^\mathrm{o}$) in the star's photosphere. The transition $4p \; ^2\mathrm{P}^\mathrm{o}$  $\rightarrow$  $3d \; ^2\mathrm{D}$ that produces the Ca II triplet emission leaves the atoms in the upper levels for the forbidden lines.  
Collisional excitation is much weaker at the relevant temperatures and 
densities.   However, some of the $3d \rightarrow 4s$ photons  
are eliminated by collisional de-excitation, whose critical electron  
density is $n_c \, \approx \, 8 \times 10^6$ cm$^{-3}$ at temperatures  
around 8000 K.  Therefore the ratio of photon fluxes  
in the two multiplets should be approximately 
   \begin{equation}
      \frac{{\Phi}({\lambda}7300)}{{\Phi}({\lambda}8600)} 
     \  \approx  \   
      \left( 1 + \frac{n_e}{n_c} \right)^{-1}  ,     
   \end{equation}
 where $n_e$ is a suitable average electron density in the emission region. 
For this calculation we used radiative transition and collision strengths for Ca II
from \citet{Melendez}. The photon ratios 
and corresponding electron densities  are given in Table 5 for the six hypergiants 
with Ca~II triplet data.
The densities range from $\sim$ 1 to 4 $\times 10^7$ cm$^{-3}$ and are comparable to our previous
results for IRC~+10420 \citep{Jones,RMH02} with $n_e$ of 2.5 $\times 10^7$ cm$^{-3}$. With reasonable mass loss rates, such densities should occur 
at $r \sim 30$ stellar radii. 

The above reasoning is subject to an obvious proviso, however.
One can easily imagine a two-component model, wherein  a denser 
region emits permitted Ca~II emission but very little forbidden
emission, while a physically distinct low-density region accounts
for nearly all of the [Ca~II] but only a fraction of the Ca~II triplet.
In that case the derived $n_e$ must be re-interpreted;  instead
of a weighted average density, $n_e$ becomes 
a lower limit for one component and an upper limit for the other.
In some of these objects a two-component model is supported by
differences between the permitted and forbidden line profiles (Fig.\ 2).
This type of model is very consistent with suspicions that
the wind is strongly non-spherical, see below.

\subsection{Thomson Scattering and Mass Loss}

The hydrogen emission profiles in the warm hypergiants  exhibit broad Thomson scattering wings, 
 and in some cases
the characteristic asymmetric profiles are also observed  in other emission lines such as the Ca II triplet and the strong Fe II emission lines.
The Thomson-scattered
wings  can help constrain the wind parameters and mass loss.  
The two best examples in this paper are M31-004444.52 and M31-004322.50 
which have  asymmetric electron 
scattering profiles  present in both the hydrogen and Fe II emission lines, see Figures 3 and 4.  On the red side of H$\alpha$ and H$\beta$ in M31-004444.52, for instance, one can 
      see a line wing at velocities of  800  km s$^{-1}$ or more from the line peak 
      (Figure 4), though the line core has a FWHM of only 200 km s$^{-1}$, and the P Cyg
            absorption implies a terminal velocity not much above 300 km s$^{-1}$.  
      Thus  the overall shape and physical context are consistent with Thomson scattering.  
      The H$\alpha$ and H$\beta$ emission profiles for the other hypergiants are shown in
      Figures 5 -- 10 in the on-line edition and their H$\alpha$ equivalant widths are included
      in Table 6. 
      (Examples of line profiles with stronger Thomson-scattered wings are  
      shown in Fig.\ 11 in \citealt{Dess08}, Fig.\ 8 in \citealt{RMH12} and several examples in hydrogen and Ca II in \citealt{RMH11}.) 
      A detailed model would be far beyond the scope of this 
      paper, but here we briefly review the case. 

Thomson-scattered line wings provide  information about circumstellar 
densities and size scales.  For these stars, however, the main 
implications are different because spherical models lead to interesting 
contradictions outlined below.
 
To illustrate the problem, consider a spherical model with 
idealized assumptions:  
(1) the outflow density is ${\rho}(r) \, \propto \,  r^{-2}$, mostly ionized;  
(2) emission line photons emitted outside radius $r_1$ escape, 
while those inside $r < r_1$ are destroyed by continuum absorption or 
by conversion to other emission lines; 
(3) the line emissivity is $A_l \, n_e^2$ where $A_l$ is a known constant 
for each line;  and 
(4) $\tau_1$, the Thomson scattering optical depth at radius $r_1$, is 
roughly indicated by the observed strength of the line wings.  When this 
type of model is used for other, more luminous objects, its main results 
agree with more elaborate calculations within factors of two or three   
(e.g., compare \citealt{RMH12,Dess08,KD95,Hillier01}). 
 It is easy to show that the above assumptions imply 
\begin{equation} 
    r_1 \ \approx \ \frac{ {\sigma}_e^2 \; L_l }
                           { 4 \, {\pi} \, {A_l} \, {\tau}_1^2 } \; ,
\end{equation}  
where $\sigma_{e}$ is the Thomson scattering cross section and 
$L_l$ is the emergent luminosity of the emission line. We apply  
this formula to H$\beta$ in M31-004444.52, for example. For  a 
 measure of the line wing, we estimate a flux ratio  
\begin{displaymath}   
 \frac { F( \mathrm{300 \ to \ 500 \ km \ s^{-1}} ) }  
 { F( \mathrm{0 \ to \ 500 \ km \ s^{-1}} ) } 
  \  \approx  \  0.1 \, ,   
\end{displaymath}  
 where the velocities are measured relative to the line peak.  
 Numerical simulations indicate that a scattering thickness   
 ${\tau}_1 \, \sim \, 0.8$ would account for this value.   
 With reasonable values $L_l \, \approx \, 4 \times 10^{36}$ erg s$^{-1}$ 
 and $A_l \, \approx \, 10^{-25}$ erg cm$^3$ s$^{-1}$ \citep{OF}, 
  eq.\ 1 then gives $r_1 \, \approx \, 30 \; R_{\odot}$.  But this is 
  absurd, because it is much smaller than the star's radius 
  $R \, \sim \, 300 \; R_{\odot}$.  If we try to improve the model by 
  allowing for local inhomogeneities in $\rho$ (``clumping''), 
  a radial dependence ${\rho}(r)$ consistent with accelerating velocities, 
  or a maximum radius for the ionization, then the deduced $r_1$ becomes 
  even smaller.  The same discrepancy occurs for the other stars in our 
  sample;   ${\tau}_1$ seems generally too large  for the observed emission 
  line fluxes.  

If this contradiction were only a factor of two or three, then we 
could ascribe it to uncertainties in the parameters;  but a 
discrepancy factor of 10 or more suggests that the line emission 
region is basically different from the above model.  
Here we consider  three possibilities: 
  \begin{enumerate}
 \item{Conceivably the line wings are not due to Thomson scattering, 
   and the true value of ${\tau}_1$ is much less than 0.8.  
   If so, however, then unexplained high velocities must be present, 
   much faster than the stars' escape velocities and the limits of 
   the observed P Cyg features. 
  In order to produce the observed fluxes in the wings, 
   these high-velocity flows would need to carry a substantial 
   fraction of the total mass loss rate.  Moreover, they 
   would most likely be non-spherical since we do not see 
  high-velocity P Cyg absorption.  (Radiative acceleration via 
   dust is not useful for this hypothesis, because most of 
  the line emission must be created much closer to the star 
   than the dust-formation radius.) }  
  \item{Perhaps we have greatly overestimated the effective 
   emissivity $A_l$, because most of the line photons may be   
   absorbed before they escape.  But this hypothesis is not 
  as simple as it may appear.  Semi-realistic models tend to 
  have outer regions where escape is easy for a freshly created 
  line photon, plus inner zones where most line photons are  
  destroyed before they can escape.  
  The {\it effective\/} rate $A_l$ is only slightly reduced in the 
  outer zones but is almost nullified in the inner zones.  
 In that case eq.\ 1 approximates the outer region, with 
  $r_1$ marking the ill-defined transition between the 
 two regimes.   Since $A_l$ thereby represents mainly the region 
  where most line photons escape, a factor-of-two alteration 
  in its effective value may be unsurprising but a factor of 
  10 seems unlikely.  Exact evaluations would depend on the 
  definition of $\tau_1$, which is too complex to discuss here. } 
 \item{Eq.\ 1 becomes invalid in a non-spherical model.  For 
 instance, if only part of the spherical shell exists, then 
 the factor $4\pi$ in the denominator is reduced so $r_1$ 
 becomes larger.  A bipolar flow may be able to achieve 
 this result, and various other clues throughout this paper 
 also suggest geometries of this type. }  
\end{enumerate}
The second and third of these possibilities allow ``reasonable'' 
mass loss rates.  For instance, consider again H$\beta$ in the case 
of M31-004444.52. Suppose that $r_1 \, \sim \, R = 300 \; R_\odot$ 
with outflows covering about 15\% of the solid angle, 
$10^{-25.3} \, < \, A_l \, < \, 10^{-25.0}$ erg cm$^3$ s$^{-1}$, 
$\tau_1 \, \sim \, 0.8$, and $V_\infty \, \approx \, 300$ km s$^{-1}$. 
We also make allowances for the effect of wind acceleration 
on $\rho(r)$.  Then the observed H$\beta$ luminosity is consistent 
with $\dot{M} / \epsilon$ in the range 
$10^{-5}$ to $10^{-4}$ $M_\odot$ y$^{-1}$ where  $\epsilon$ is 
a standard volume-filling factor to allow for local inhomogeneities, 
$0 < \epsilon < 1$. (A mass loss range is estimated for M31-004444.52 and M31-004322.50 in 
Table 6 if
$\epsilon$ is $\sim$ 1.)  A more detailed discussion would be very lengthy 
and is beyond the scope of this paper, but the main point is this:  
the main parameters are reasonable {\it if we do not assume spherical 
symmetry.\/}

Bipolar outflows can  explain why some of these objects show
strong P Cyg absorption while others do not;  why some but not
all of them have asymmetric line profiles;  and why double line
peaks occur in some cases.  The forbidden [Ca~II] emission noted
in {\S}4.1 may originate in lower-density low-latitude zones
of the outflow.  Thus, although we cannot {\it prove\/} that
these objects have polar outflows, the concept is very appealing.

The line profiles also imply that the winds are
 not opaque in the continuum; in other words the photospheres
do represent stellar surfaces and not dense outflows.
Based on ``thermalization depth'' arguments, the color-temperature
photosphere in an opaque wind usually corresponds to a
Thomson-scattering optical depth of the order of 3
\citep{KD87}.  Therefore an opaque wind
should be accompanied by Thomson-scattered emission line wings
with $\tau_1 \sim 2$, appreciably larger than the $\tau_1 < 1$
estimated above.

\subsection{Circumstellar Dust  and Mass Loss}

The SEDs of almost all of the stars described here show the presence of circumstellar dust
in addition to the gaseous ejecta. The SEDs are shown in Figures 11 and 12 for the M31 and M33 stars, respectively. Var A is shown separately in Figure 13 reproduced from \citep{RMH06}
with the more recent  photometry from WISE and its current V magnitude (Table 3). 
The one exception is B324 shown in Figure 14.  Its interstellar extinction corrected
SED instead shows  significant near-infrared radiation due to strong free-free emission  from its wind. 
We attribute its  apparent  excess longwards of 10$\mu$m from the WISE data
to PAH and dust emission from an H~II region.
B324 is in a dense association (No. 67 in \citealt{HS80}) of luminous, young stars 
and their associated nebulosity. The H$\alpha$ image from \citet{Massey06} shows B324 embedded in emission 
nebulosity.   

The SEDs of the remaining warm hypergiants show thermal emission from dust to varying 
degrees in the near- and mid-infrared from 3.5$\mu$m to 8$\mu$m in the IRAC data and
in most cases the WISE data  shows that the thermal emission also extends to even longer 
wavelengths.
We can use the flux at mid-infrared wavelengths to estimate the mass of the 
dusty circumstellar material with some assumptions about the grains.

The dust mass is given by  

 \[  
        M_{dust} = \frac{4D^{2} {\rho} {\lambda}F_{\lambda} }
	             { 3({\lambda} Q_{\lambda}/a) B_{\lambda}(T) } ,  
		                \]

where D is the distance to M31 or M33, F$_{\lambda}$ is the mid-infrared flux, {\it a} is the grain 
radius, $\rho$ $\approx$ 3 g cm$^{-3}$ is the grain density, Q$_{\lambda}$ is the grain efficiency
for the absorption and emission of radiation, and B$_{\lambda}$(T) is the Planck specific intensity at temperature T. 

The SEDs of the warm hypergiants are relatively flat suggesting that the dust grains are 
radiating at a range of temperatures from possibly as warm as 1000 K to 300 K at 10$\mu$m and in
an extended zone from $\sim$ 100 AU to several 100 AU from the star. Except for Var A, we do not have a 
measurement of the 9.8 and 18$\mu$m silicate features. We therefore adopt the flux at 8$\mu$m 
for the other stars and an assumed grain temperature of 350 K for this calculation using the formulation from 
\citet{Suh}  for Q$_{\lambda}$ with $a = 0.1\mu$m. Assuming the nominal gas 
to dust ratio of 100 we then derive estimates of the total mass lost in Table 6. The results show a range of at least a factor of 10 for the mass of the circumstellar material, although most of the hypergiants have 
apparently shed about 2 $\times$ 10$^{-2}$ M$_{\odot}$. 

For Var A, we use the flux in the 9.8$\mu$m silicate feature to estimate the mass loss and its mass loss rate   
 at two different times, during its high mass loss event (1986) and after its cessation had begun (2004), see Figure 7. Not surprisingly these give different mass 
loss estimates.  Assuming that the dust formed as a result of the high mass loss event in 1951,
we find a mass loss rate during the event of 5 $\times$ 10$^{-4}$  M$_{\odot}$ yr$^{-1}$   which  declined to an average of 
1.5 $\times$ 10$^{-4}$  M$_{\odot}$ yr$^{-1}$ after Var A  had transitioned back to its warmer state. The former value agrees with our earlier result \citep{RMH87} using an independent method. 

We also determine the total or bolometric luminosities of the hypergiants by integrating
their SEDs from 0.35$\mu$m to 10$\mu$m, first  corrected for interstellar extinction. 
There is always some uncertainty
associated with visual extinction estimates from the observed colors for stars with 
 emission lines in their spectra. We therefore estimate the extinction the 
classical way from the observed $B-V$ color and apparent spectral type, and also for 
comparison, from nearby stars  assuming that their 
UBV colors from \citet{Massey06} are normal. We adopt R$=$ 3.2 mag for the ratio of 
total to selective extinction. For example,
the observed $B-V$ colors for N093351 would imply essentially zero reddening for its
apparent spectral type even though the foreground A$_{v}$ for M33 is $\approx$ 0.3 mag.
Two nearby stars, including UIT231 \citep{Massey96} plus a fainter OB star, give A$_{v}$
of 0.5 and 0.4 mag. We therefore adopt 0.45 mag for N093351. A similar situation occurs for 
M31-004522.58. For three of the other stars,  however the extinction estimate from the colors 
was systematically higher than from the neighboring stars (Table 7). The only star for which the
two A$_{v}$ estimates are comparable, B324,  has little or no circumstellar dust, thus the higher A$_{v}$ for the other stars from their colors 
might be due to additional circumstellar reddening.  
 The higher extinction values also produced SEDs in the 
visual more consistent with the temperatures  expected from their spectra. We therefore adopted
A$_{v}$ from the observed colors.  
For purposes of comparing
stars in different galaxies we use the Cepheid distance scale with moduli,  
  24.4 mag  for M31 \citep{M31Ceph} and 24.5 mag for M33 \citep{M33Ceph} for the luminosities in Table 7.

B324 is the most luminous with an absolute 
bolometric magnitude of -10.1 and a luminosity of $\approx$ 8 $\times$ 10$^{5}$ L$_{\odot}$ using
this distance for M33 and the visual magnitudes measured from the  HST images (Table 3).   
At a temperature of $\sim$ 8000 K, it is not above the empirical upper luminosity boundary. Instead, as one of the visually most luminous stars in M33, B324  is one of the stars that defines it for that galaxy.  

Var A's luminosity in Table 7 may appear anomalously low. Based on its visual maximum in 1951 and
its infrared flux in 1986 \citep{RMH87}, its absolute bolometric magnitude was $\approx$ -9.5 (5 $\times$ 10$^{5}$ L$_{\odot}$). We  \citep{RMH06} showed 
that after about 45 yrs, Var A's  high mass loss event had 
ceased.  Its spectrum and colors had returned to its F  supergiant state, although  it remained 
obscured, and the 10$\mu$m flux  and corresponding mass loss rate (see above) had  declined.  
 As we emphasized in 2006, a decrease in 
total luminosity  by  a factor of 3 -- 5  is unrealistic.  Consequently  we  concluded that 
much of the radiation must be escaping  from out of our line of sight. With the recent evidence that Var A may be starting to brighten in the visual, continued observations may provide clues to 
the structure of its circumstellar ejecta and mass loss as it recovers from its  high mass loss episode.

Except for M31-004322.50, which is significantly fainter than the other stars,  the warm 
hypergiants have luminosities that suggest they had initial masses above 20 M$_{\odot}$ with 
some of them likely above 30 M$_{\odot}$.  

\section{Comments on the Evolutionary State of the  Warm Hypergiants}
 
It has been acknowledged for some time that the Galactic hypergiant IRC~+10420 is a post-red supergiant \citep{Jones,Rene96}.   Var A and $\rho$ Cas \citep{deJ98}, with their apparent transits in the HR Diagram,
are likely in a similar unstable
post-RSG state. In this paper we are suggesting that these seven stars in M31 and M33, including Var A, 
are candidates for 
post-red supergiant evolution. They have the characteristics we would expect; A to F-type absorption spectra,  winds with relatively slow outflows, an extensive and dusty circumstellar ejecta, and relatively high mass loss. B324 may be the 
one exception. Although it has a low density circumstellar envelope, it lacks a dusty environment, but then so do
$\rho$ Cas and HR 8752. Its evolutionary state is thus uncertain. It could just as easily be evolving towards
cooler temperatures with significant mass loss due to  its high luminosity or be on a blue loop as a post-RSG.

Some authors have previously suggested that some of these stars are LBVs, presumably in their maximun light 
phase when their optically thick winds  resemble A to F-type supergiants. If these stars were above the empirical luminosity boundary in the HR Diagram,
then it would be easy to argue that they are LBVs in their  ``eruption'' state. But the
warm hypergiants have luminosities that place them below the upper luminosity boundary. On the
HR Diagram these  warm hypergiants will lie between $\approx$ 7000 and 9000K which does indeed overlap the maximum light 
phase for the LBVs (Figure 9  in \citet{HD94}).  However unlike the LBVs,  the winds of these warm hypergiants are not optically thick based on the discussion in \S {4.2}. 

Furthermore, the lack of significant 
variability over several decades for the M33 stars (Paper III) and no evidence of large brightness variation in the three M31 candidates,  support a post-RSG origin as opposed to LBVs in an extended maximum. The warm hypergiants do show the small oscillations  ($\pm$ 0.1 -- 0.2 mag) typical of A to F-type supergiants, often referred to as $\alpha$ Cygni variability (see \citet{vanG2002} and references therein). LBVs at maximum light also exhibit  $\alpha$ Cygni variability, but during an 
extended maximum show larger variations of $\pm$0.5 mag or more (van Genderen et al. 1997a,b).
The warm hypergiants, however,  are very likely the progenitors of the ``less luminous'' LBVs 
which have presumably been red supergiants \citep{HD94}. They could indeed be the progenitors of a star like 
R71 or even SN 1987A.

We conclude by emphasizing that this is by no means a comprehensive survey for post-RSGs. These stars were identified as
part of a larger program on the luminous and variable stars in M31 and M33 (Paper II). It is also possible that not all
post-RSGs will share all of the characteristics of this small subset. Furthermore, as they evolve to warmer temperatures on their
blue loops they may become harder to distinguish from the normal supergiants unless they do become LBVs or 
develop distinctive  emission-line spectra perhaps like the Fe II emission-line stars with large 
infrared excesses discussed in our next paper.

\acknowledgements
Research by R. Humphreys and K. Davidson on massive stars is supported by  
the National Science Foundation grant AST-1019394. J.C. Martin's collaborative work on luminous variables is supported by the National Science Foundation grant  AST--1108890.  
This paper uses data from the MODS1 spectrograph built with funding from NSF grant AST-9987045 and the NSF Telescope System Instrumentation Program (TSIP), with additional funds from the Ohio Board of Regents and the Ohio State University Office of Research.
This publication also makes use of data products from the Wide-field Infrared Survey Explorer, which is a joint project of the University of California, Los Angeles, and the Jet Propulsion Laboratory/California Institute of Technology, funded by the National Aeronautics and Space Administration.

{\it Facilities:} \facility{MMT/Hectospec, LBT/MODS1}

\clearpage


\input{Table1.tex}

\clearpage

\input{Table2.tex}

\clearpage

\input{Table3.tex}

\clearpage

\input{Table4.tex}

\input{Table5.tex}

\input{Table6.tex}

\input{Table7.tex}

\clearpage 


\begin{figure}
\figurenum{1}
\epsscale{1.0}
\plotone{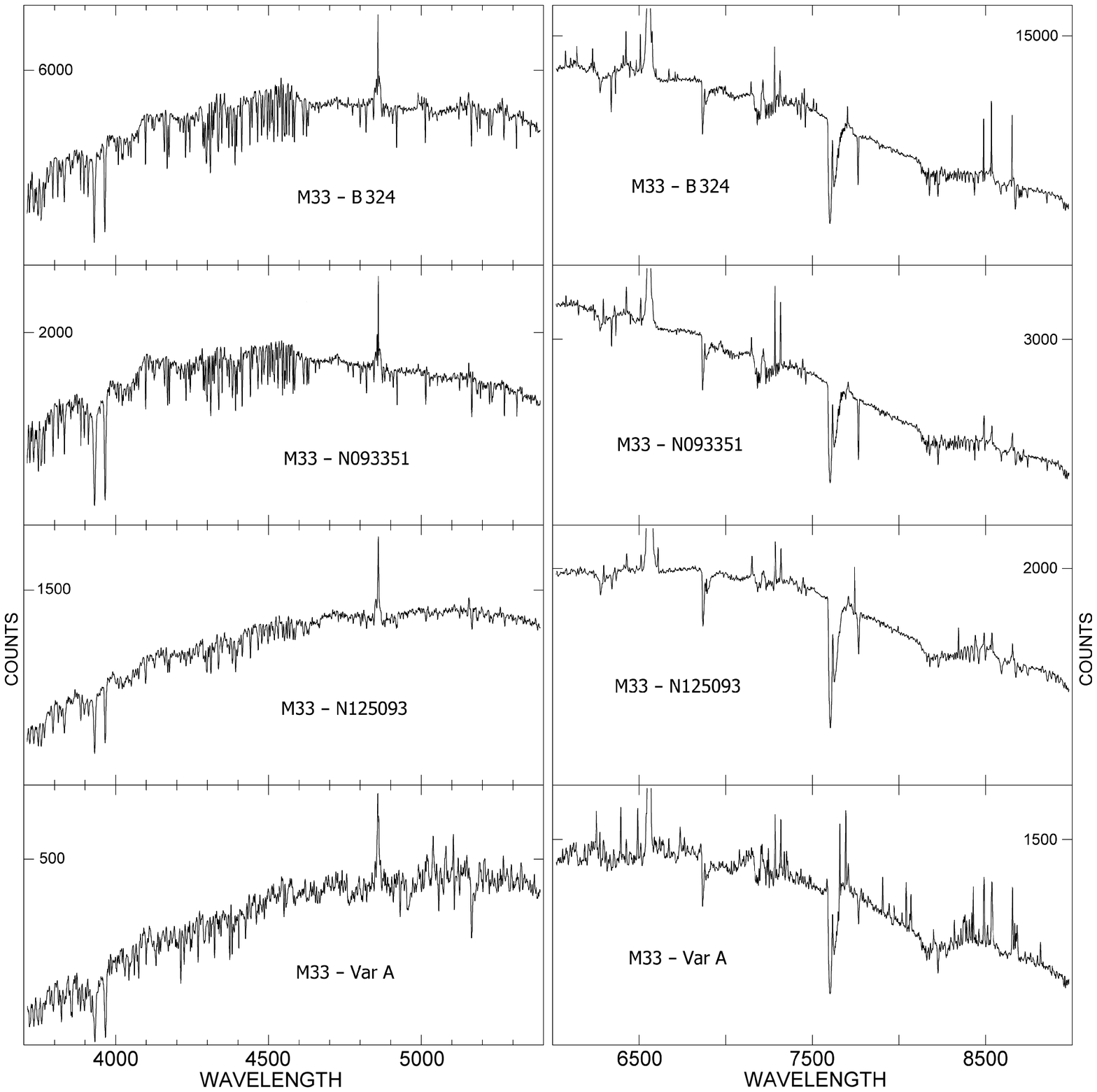}
\caption{The blue spectra and red spectra of the warm hypergiants in M33. All spectra are from the LBT/MODS spectrograph.}
\end{figure}

\begin{figure}
\figurenum{2}
\epsscale{0.8}
\plotone{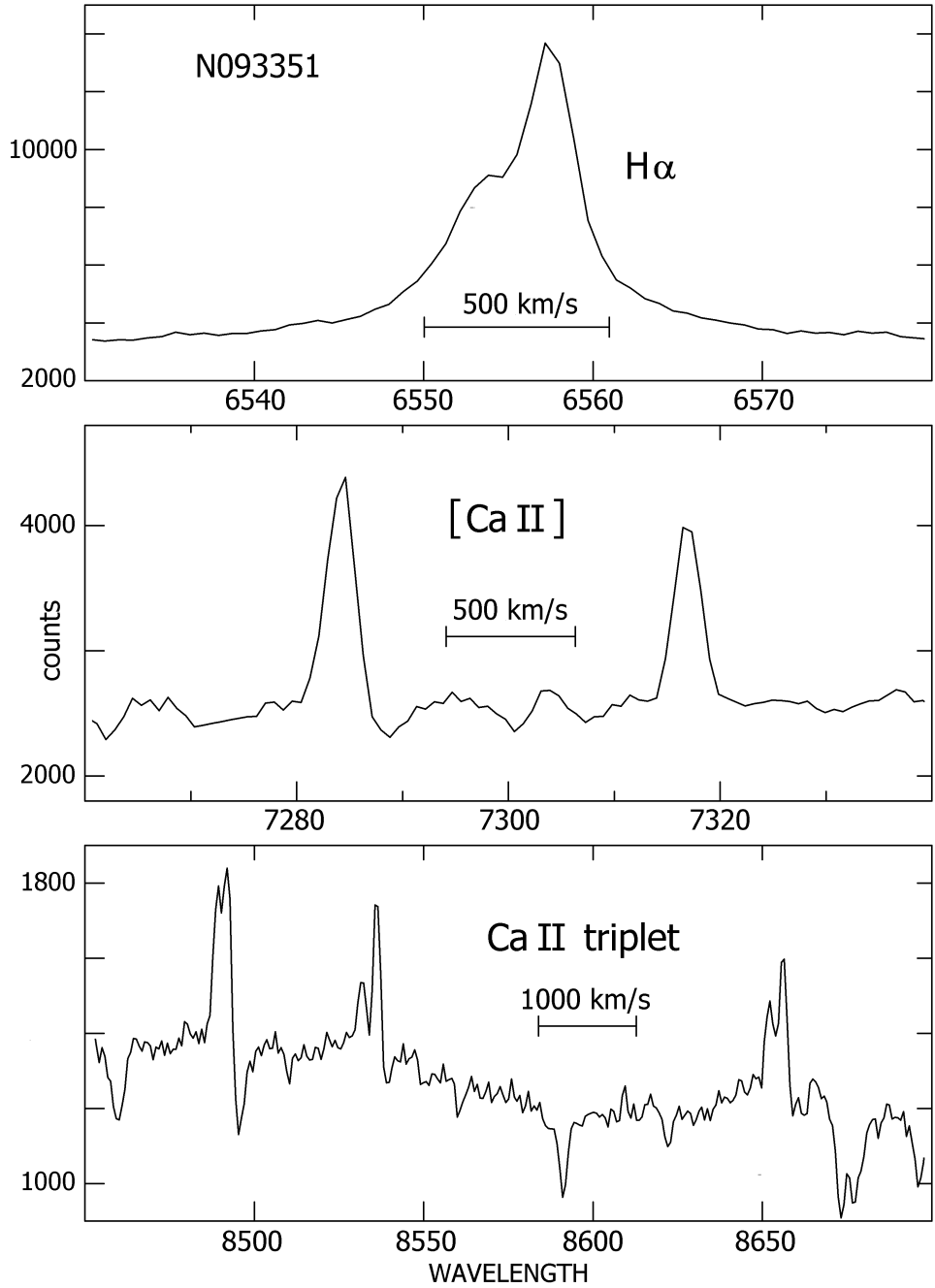}  
\caption{The H$\alpha$, [Ca II] and Ca II triplet emission profiles in N093351. The double Ca II 
emission is not due to a blend with emission in the Paschen lines. Note that the Paschen line 
at $\lambda$8598{\AA} is in absorption.}
\end{figure}

\begin{figure}
\figurenum{3}
\epsscale{1.0}
\plotone{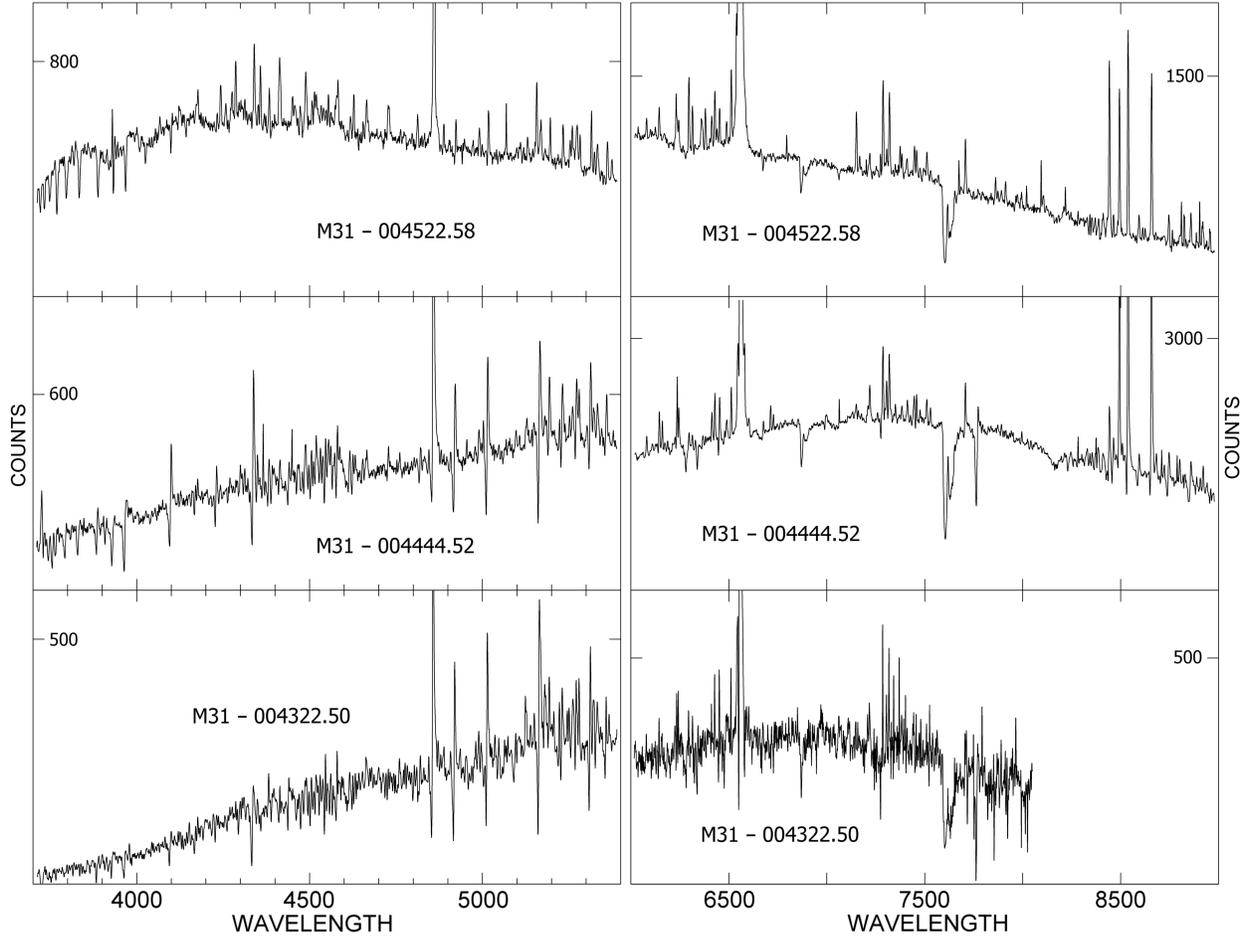}
\caption{The blue spectra and red spectra of the warm hypergiants in M31. The spectra of M31-004322.50 are from the MMT/Hectospec and do not inlcude the Ca II triplet in the far red.}
\end{figure}

\begin{figure}
\figurenum{4}
\epsscale{0.8}
\plotone{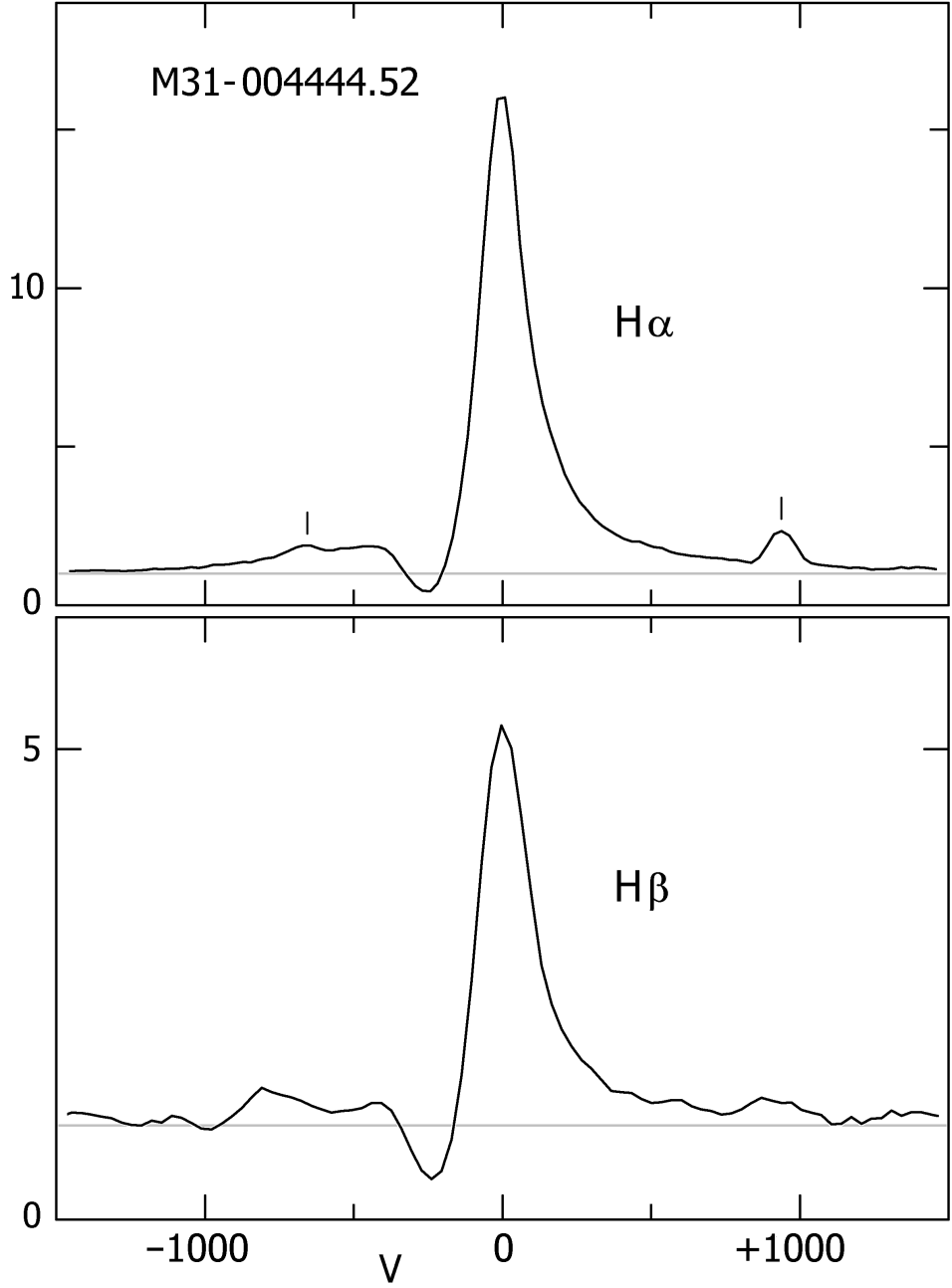}
\caption{The asymmetric H$\alpha$ and  H$\beta$ profiles for M31-004444.52. The tic marks identify the [N II] lines  due to nebular emision.} 
\end{figure}

\begin{figure}
\figurenum{11}
\epsscale{0.6}
\plotone{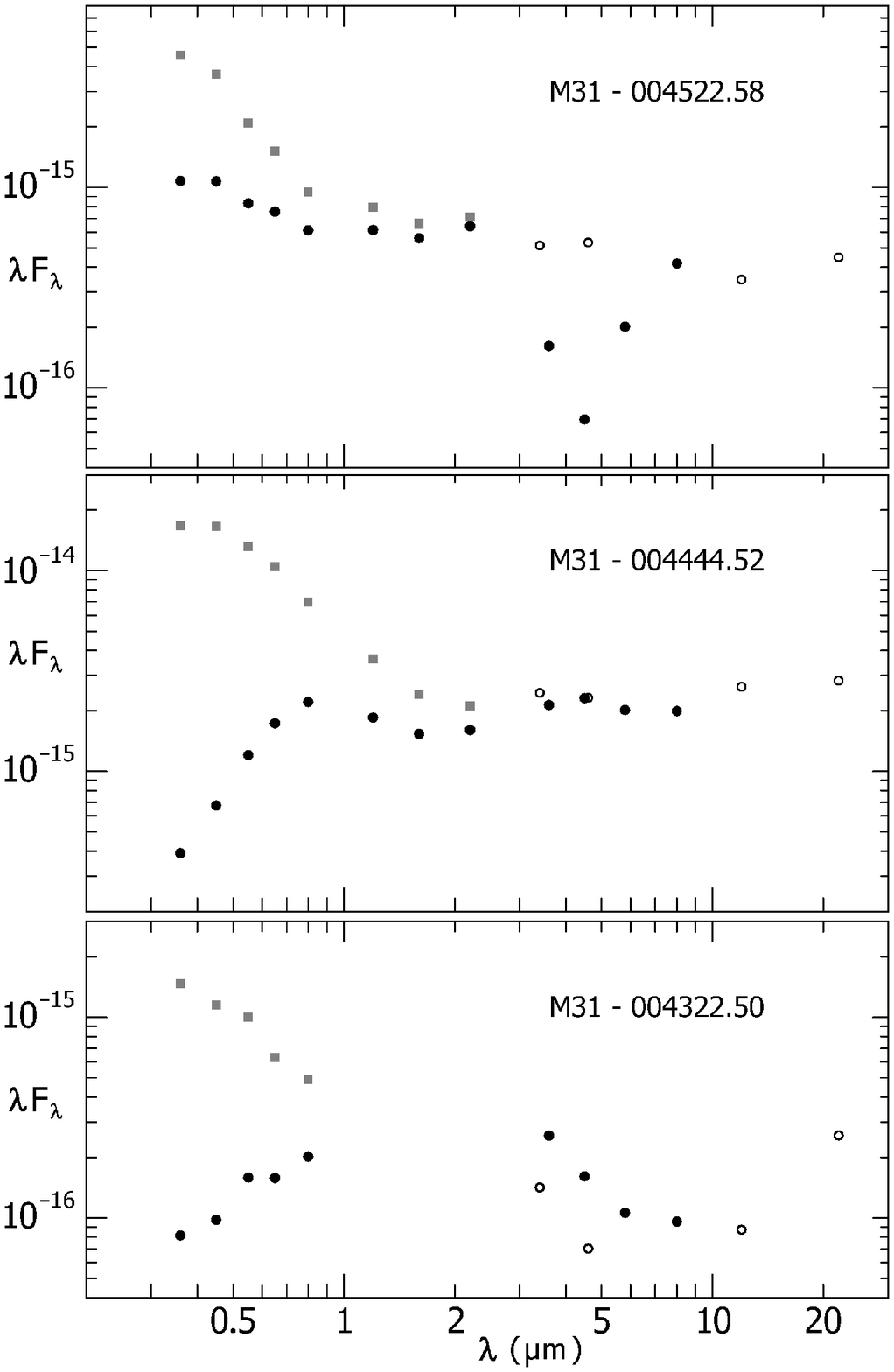}
\caption{The SEDs for the three warm hypergiants in M31. The units are Watts/m$^{2}$ vs. wavelenght in microns. The observations are plotted as filled circles with the WISE data as open circles, and 
the interstellar extinction corrected fluxes are shown as squares. The mid-infrared Spitzer and WISE photometry for M31-004522.58 are obviously discrepant. The prominent dip in the Spitzer data 
is the chracteristic signature of an H II region; however its spectrum does not show any evidence for nebular emision and the longer wavelength WISE data does not show the rise in flux expected from an H II region. The upturn in the SED for 
M31-004322.50 at 20$\mu$m is most likely due to nebular emission.}
\end{figure}

\begin{figure}
\figurenum{12}
\epsscale{0.8}
\plotone{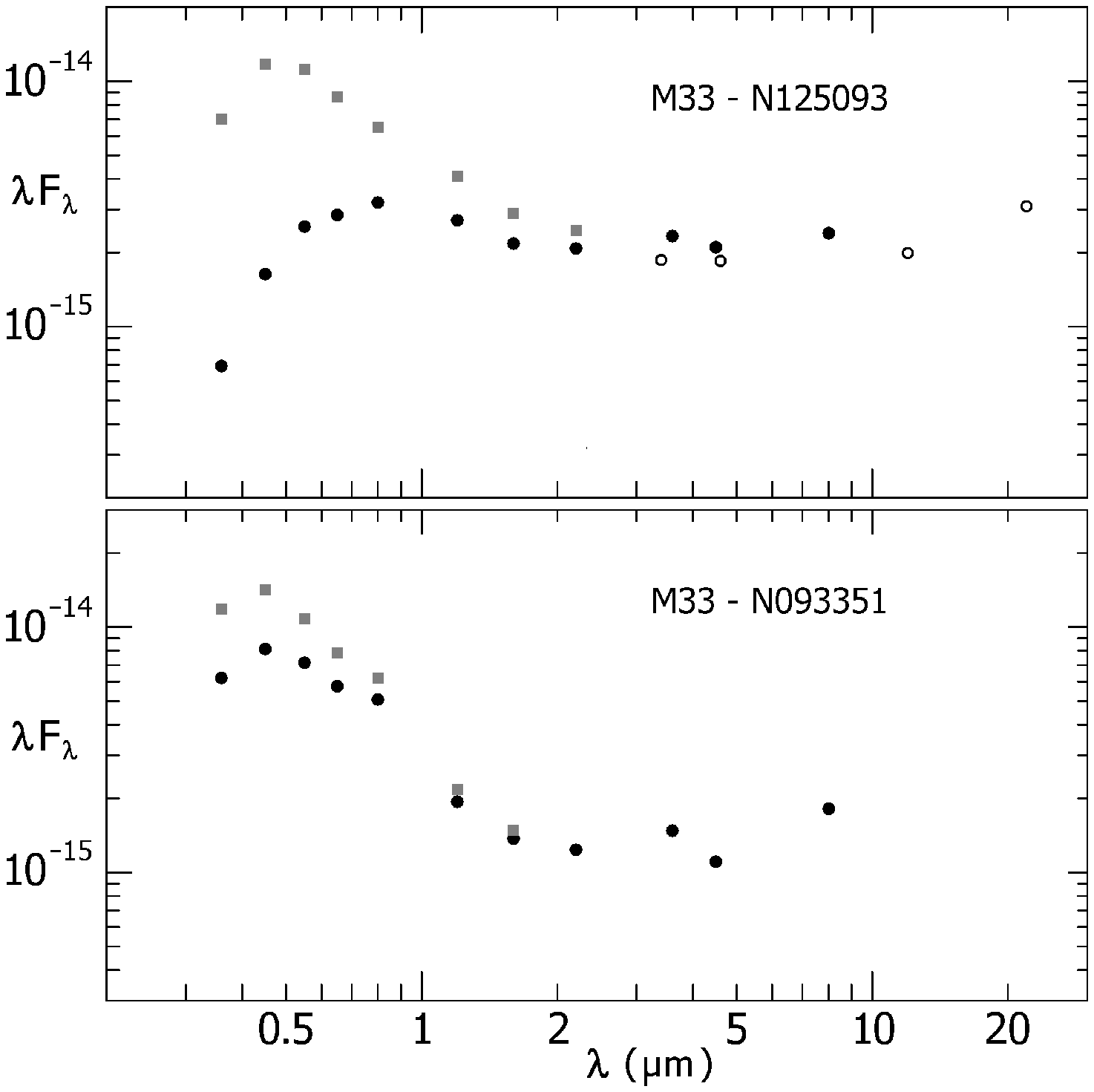}
\caption{The SEDs for two warm hypergiants in M33. The units and symbols are the same as in Figure 11}
\end{figure}

\begin{figure}
\figurenum{13}
\epsscale{1.0}
\plotone{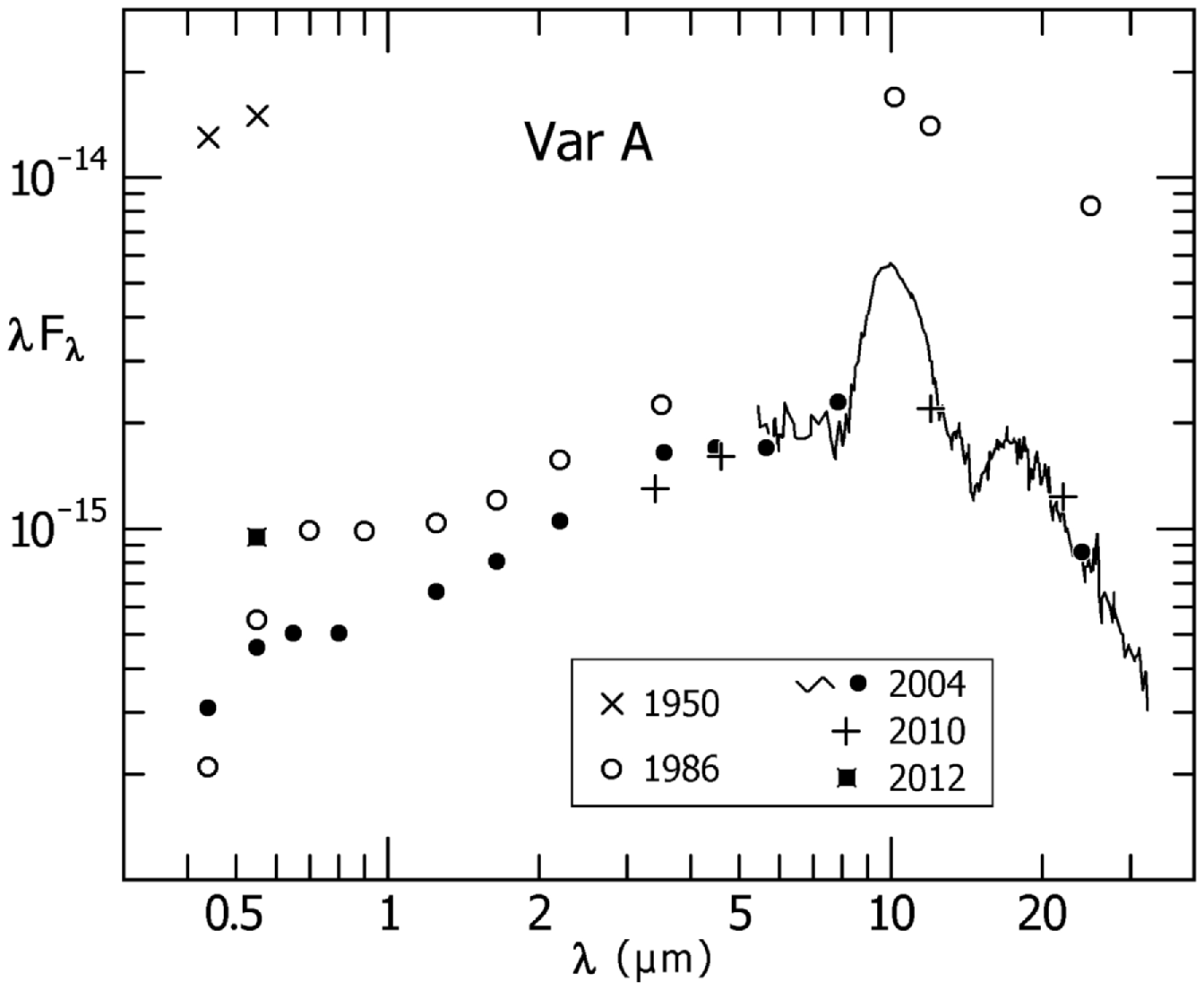}
\caption{The SED for Var A reproduced from \citet{RMH06} with addition of the WISE data ( +'s) and 
the recent visual point plotted as a square (Table 3). The vertical scale is in Watts/m$^{2}$ for comparison with the other hypergiants.}
\end{figure}

\begin{figure}
\figurenum{14}
\epsscale{1.0}
\plotone{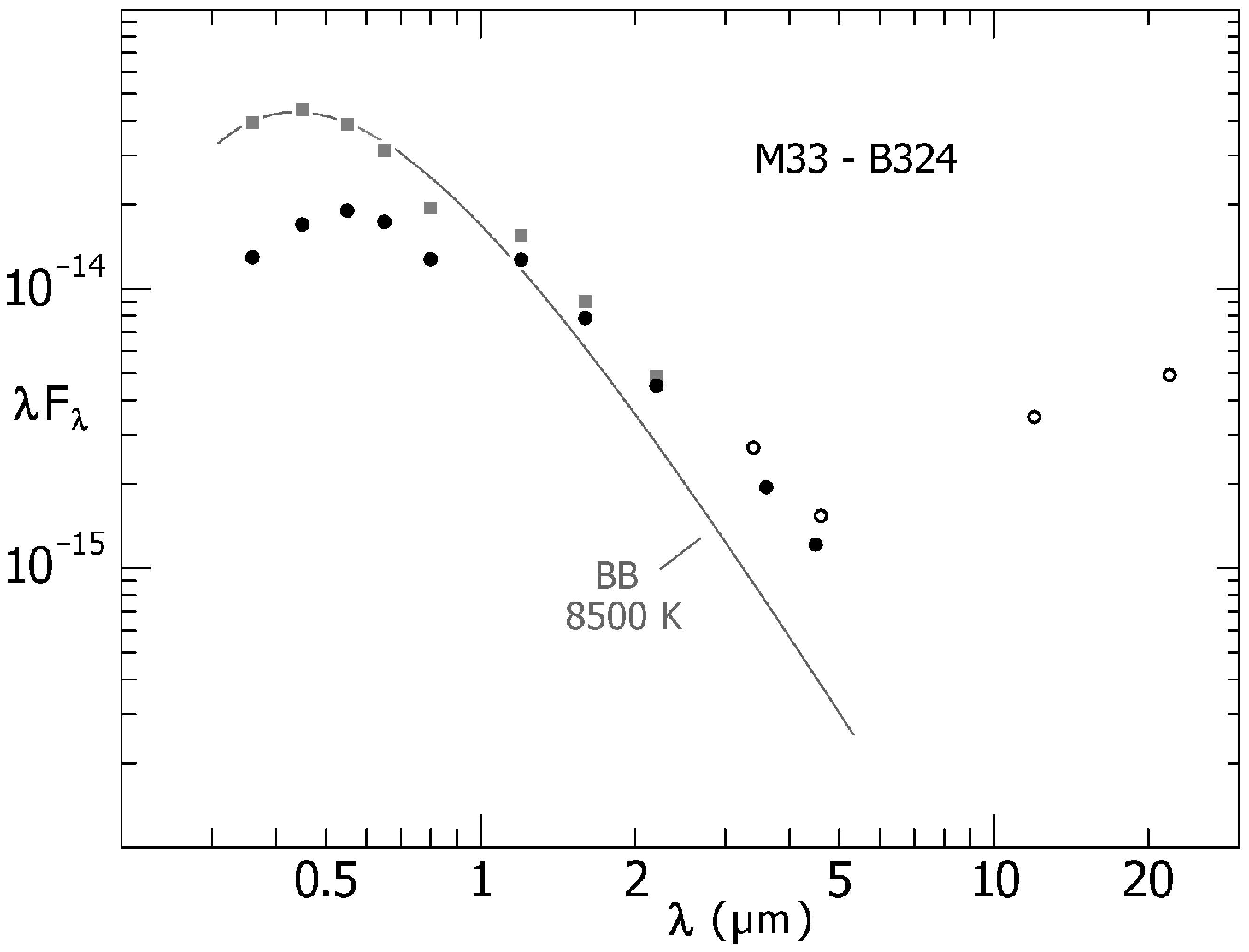}
\caption{The SED for B324. The units and symbols are the same as in Figures 11 and 12. A 8500 K blackbody is shown fit through the extinction-corrected visual photometry, and illustrates that the
near-IR excess is due to free-free emission. The increased flux longwards of 10$\mu$m in the WISE data is most likely due to emission from the surrounding nebulosity.}
\end{figure}







\end{document}

%% file: Table1.tex
\begin{deluxetable}{lllllcl}
\tablewidth{0 pt}
\tabletypesize{\footnotesize}
\tablenum{1} 
\tablecaption{Journal of Observations}
\tablehead{
\colhead{Target} &
\colhead{UT Date} &
\colhead{Spectrograph}  &
\colhead{Exp. Time} &
\colhead{Grating \& Tilt} &
\colhead{Slit/Aperture($\arcsec$)} & 
\colhead{Seeing($\arcsec$)} 
}
\startdata 
M31-Blue  &   10 Oct. 2010  &  MMT/Hectospec & 120m  &  600l, 4800{\AA}& 1.5 & 1.0 \\
M31-Red   &   09 Oct. 2010  &  MMT/Hectospec &  90m  &  600l, 6800{\AA}& 1.5 & 1.0\\
M33-Blue  &   03 Oct. 2010  & MMT/Hectospec & 120m  &  600l, 4800{\AA} & 1.5 & 0.4 \\
M33-Red   &  03 Oct. 2010  & MMT/Hectospec &  90m  &  600l, 6800{\AA} &  1.5 & 0.4 \\
M33-VarA  &  26 Sept. 2011 & LBT/MODS1   &  21m   & Dichroic& 0.6 &  0.8 - 1.0  \\       
M33-B324  &  26 Sept. 2011 & LBT/MODS1   &  15m   & Dichroic& 0.6 &  1.0 - 1.4\\
M33-N093351 &  28 Sept. 2011 & LBT/MODS1   &  6m   & Dichroic & 0.6 &  0.5  \\
M33C-4119   & 29 Sept. 2011 & LBT/MODS1   &  12m  & Dichroic  & 0.6 &  0.6 \\ 
M31-004417.10 & 13 Oct. 2012 & LBT/MODS1   & 12m & Dichroic & 1.0 &  1.2 \\
M31-004522.58 & 13 Oct. 2012 & LBT/MODS1   & 18m & Dichroic & 1.0 & 1.7  \\
M33-N045901   & 14 Oct. 2012 & LBT/MODS1   & 15m & Dichroic & 1.0 &  0.9 \\
M33-N125093 & 14 Oct. 2012 & LBT/MODS1   & 15m & Dichroic & 1.0 &  0.9 \\
M33C-15731  & 14 Oct. 2012 & LBT/MODS1   & 12m & Dichroic  & 1.0 &  0.7  \\ 
M31-004425.18 & 17 Nov. 2012 & LBT/MODS1 & 15m & Dichroic & 1.0 &  1.3 - 1.6 \\
M31-004444.52 & 17 Nov. 2012 & LBT/MODS1 & 18m & Dichroic & 1.0 &  1.2 - 1.5\\
M33C-7292     & 17 Nov. 2012 & LBT/MODS1 & 12m & Dichroic  & 1.0 &   1.0 - 1.2\\
M31-004229.87 & 05 Jan. 2013 & LBT/MODS1 & 23m & Dichroic & 1.0 &  0.8 \\ 

\enddata
\end{deluxetable}

%% file: Table2.tex
\begin{deluxetable}{lcccl}
\tablewidth{0 pt}
\tabletypesize{\footnotesize}
\tablenum{2} 
\tablecaption{The Warm Hypergiants  in M31 and M33}
\tablehead{
\colhead{Star Name} &
\colhead{RA (2000)} &
\colhead{Dec (2000)}  &
\colhead{Spec. Type} &
\colhead{Other Id/Notes/References}  
}
\startdata 
   &    &   M31   &    &    \\
M31-004322.50  &  J004322.50   &  +413940.9  &   late A - F0  I &   \\
M31-004444.52 &   J004444.52  &   +412804.0   &  F0 Ia  &    \\
M31-004522.58 &   J004522.58  &  +415034.8    &  A2 Ia  &    \\ 
   &               &              &                               \\
   &               &   M33        &                                \\
Var A & J013232.80 & +303025.0    &  F8 Ia & Humphreys et al. (1987,2006) \\
N093351  &     J013352.42 &  +303909.6 &  F0 Ia & \\  
B324  &  J013355.96 &  +304530.6 &   A8-F0 Ia  &  UIT 247 \\
N125093  & J013415.38  & +302816.3  &  F0-F2 Ia & \\ 
 
\enddata
\end{deluxetable}

%% file: Table3.tex
\begin{deluxetable}{lllllllllllllllll}
\rotate
\tablewidth{0 pt}
\tabletypesize{\scriptsize}
\tablenum{3} 
\tablecaption{Multi-Wavelength Photometry (in magnitudes)}
\tablehead{
\colhead{Star} & 
\colhead{U}  &
\colhead{B} &
\colhead{V} &
\colhead{R} & 
\colhead{I} &
\colhead{J} &
\colhead{H} &
\colhead{K} &
\colhead{3.6$\mu$m}\tablenotemark{a} &
\colhead{4.5$\mu$m}\tablenotemark{a} &
\colhead{5.8$\mu$m}\tablenotemark{a} &
\colhead{8$\mu$m}\tablenotemark{a}  &
\colhead{3.4$\mu$m}\tablenotemark{b} &
\colhead{4.6$\mu$m}\tablenotemark{b} &
\colhead{12$\mu$m}\tablenotemark{b} &
\colhead{22$\mu$m}\tablenotemark{b}
}

\startdata
M31-004322.50  & 20.7& 	21.2	& 20.3	& 19.9	& 19.2 & \nodata & \nodata & \nodata & 14.9& 	14.6& 	14.4  & 13.5 & 15.7&	15.5&	12.3&	9.1 \\   
M31-004444.52  &  19	& 19.1	& 18.1	& 17.3	& 16.6 & 15.8&	15.2&	14.4 & 12.6	& 11.8	& 11.2	& 10.2 & 12.6	&11.7	&8.6	& 6.5 \\   
M31-004522.58 & 17.9&	18.6& 	18.5& 	18.2	& 18 & 17& 	16.3& 	15.4 & 15.4	& 15.6	&  13.7	&  11.9  & 14.2	&13.3	&10.8	&8.5 \\  
M33-Var A\tablenotemark{c}& 20.2: & 19.9  & 19.1  & 18.6 & \nodata  & 16.9  & 15.9 &  14.7 & 12.9   & 12.1 & 11.4  &  10.1  & 13.30& 	12.10&	8.80	& 7.40 \\
M33-N093351 &  16& 	16.4& 	16.2&	16& 	15.7 & 15.7&	15.3& 	14.7 & 13.0 & 12.6 & \nodata & 10.3 & \nodata & \nodata & \nodata & \nodata \\ 
M33-B324\tablenotemark{d} & 14.9 & 15.3	& 14.9	& 15.2	& 15.0 & 13.7& 	13.4& 	13.3 & 12.7&	12.5  & \nodata & \nodata &  12.50&	12.15&  	8.30& 	5.90 \\  
M33-N125093 & 18.4&	18.1&	17.3& 	16.8 &	16.2 & 15.4&	14.8&	14.1 & 12.5& 	11.9&  \nodata  & 10.0 & 12.90	& 11.90	 & 8.90	& 6.40 \\

\enddata
\tablenotetext{a}{{\it Spitzer}/IRAC}
\tablenotetext{b}{WISE}
\tablenotetext{c}{The photometry is from \citet{RMH06}. Recent CCD photometry
obtained at the Barber Observatory, Univeristy of Illinois Springfield, in October and December 2012 suggests that 
Var A may have begun to brighten.  Its current  V magnitude is 18.36 $\pm$ 0.05 
 measured relative to more than 40 comparison stars in the same field. }
\tablenotetext{d}{B324 is in a very crowded region and groundbased visual 
photometry of B324 can be  contaminated by nearby faint stars withn an arcsec
or so of B324. For that reason we measured magnitudes from HST/WFPC2 F555W and F439W obtained November 30, 1998 which when converted to the standard V and B magnitudes give 15.12 and 15.56 respctively, slightly fainter than the groundbased photometry. }
\end{deluxetable}

%% file: Table4.tex
\begin{deluxetable}{lccccc}
\tablewidth{0 pt}
\tabletypesize{\footnotesize}
\tablenum{4} 
\tablecaption{Outflow Velocities (km s$^{-1}$) }
\tablehead{
\colhead{Star} &
\colhead{Double H$\alpha$} & 
\colhead{Double Ca II}  &
\colhead{P Cyg (H)} &
\colhead{P Cyg (Ca II/[Ca II])} &
\colhead{P Cyg (Fe II)} 
}
\startdata 
M31-004322.50  & \nodata   & \nodata   & -326   & -90   & -200 \\
M31-004444.52 &  \nodata  &   \nodata   & -320  & -254 & -301  \\
M31-004522.58 &  $\pm$ 100  &  \nodata    &  \nodata  & \nodata & \nodata  \\ 
N093351  &   $\pm$ 120 & $\pm$ 70  &  -92.5 &  \nodata  & \nodata\\  
B324  &  \nodata  & \nodata  &  -143  &  -126  &  \nodata \\
N125093  & $\pm$ 220   & \nodata  &  -160  & \nodata &  \nodata \\ 
\enddata
\end{deluxetable}

%% file: Table5.tex
\begin{deluxetable}{lcc}
\tablewidth{0 pt}
\tabletypesize{\footnotesize}
\tablenum{5} 
\tablenum{5}  
\tablecaption{Circumstellar Densities based on [Ca~II] and Ca~II } 
\tablehead{
\colhead{Star} &
\colhead{Photon ratio\tablenotemark{a}}  &
\colhead{ $n_{e}/(10^{7}$ cm$^{-3})$\tablenotemark{b} }  }     
\startdata 
M31-004444.52 &  0.16    &  4.2  \\
M31-004522.58 &  0.20    &  3.2  \\ 
Var A         &  0.24    &  2.5  \\
N093351       &  0.50    &  0.8  \\  
B324          &  0.37    &  1.4  \\
N125093       &  0.36    &  1.4  \\ 
\enddata  
\tablenotetext{a} { Ratio of multiplets,  
     $\Phi(\lambda7300)/\Phi(\lambda8600)$. }  
     \tablenotetext{b} { Assuming $n_c = 8 \times 10^6$ cm$^{-3}$ for 
          collisional de-excitation of [Ca~II] $\lambda7300$. } 
\end{deluxetable}

%% file: Table6.tex
\begin{deluxetable}{llcc}
\tablewidth{0 pt}
\tabletypesize{\footnotesize}
\tablenum{6} 
\tablecaption{Mass Loss Estimates}
\tablehead{
\colhead{Star} &  W$_{\lambda}($H$\alpha$)\tablenotemark{a}  &
\colhead{Mass Loss Rate (M$_{\odot}$) yr$^{-1}$} &
\colhead{Mass Lost (M$_{\odot}$)} \\  
  & {\AA}  &   Thomson scattering  &  IR (SED)   
}
\startdata 
M31-004322.50  &  96 & $\sim$  10$^{-6}$ to 10$^{-5}$  & 0.09 $\times$ 10$^{-2}$      \\
M31-004444.52 &   91  &  10$^{-5}$ to  10$^{-4}$    & 1.83 $\times$ 10$^{-2}$  \\
M31-004522.58 &   171 (LBT) & \nodata  & 0.37 $\times$ 10$^{-2}$     \\ 
Var A   &  43  & 5 $\times$ 10$^{-4}$, see text  &  2.40 to 0.8 $\times$ 10$^{-2}$\tablenotemark{b}     \\
N093351  &  29 (LBT) &  \nodata  & 1.83 $\times$ 10$^{-2}$     \\  
B324     &  34 (LBT)  & \nodata     &  no dust  \\
N125093  &  53  & \nodata  & 2.40 $\times$ 10$^{-2}$    \\ 
\enddata
\tablenotetext{a}{The hydrogen emission lines in some of the hypergiants are contaminated with nebular emission. This is most obvious in the MMT spectra for B324 and N093351. We use 
their LBT spectra obtained with a narrower slit, see text and Figure 5 in the on-line edition.} 
\tablenotetext{b}{The mass lost from Var A is estimated from its  flux in the 9.8$\mu$m silicate feature at two different times during its high mass loss episode, 1986 and 2004, see text.}
\end{deluxetable}

%% file: Table7.tex
\begin{deluxetable}{lcccc}
\tablewidth{0 pt}
\tabletypesize{\footnotesize}
\tablenum{7} 
\tablecaption{Visual and Total Luminosities }
\tablehead{
\colhead{Star} &
\colhead{A$_{v}$ (colors)} &
\colhead{A$_{v}$ (stars)}  &
\colhead{M$_{v}$} &
\colhead{M$_{Bol}$} \\
    &   (mag)  & (mag)  &  (mag)  & (mag) 
}
\startdata 
M31-004322.50  & 2.0  & 1.5  & -6.1  & -6.2    \\
M31-004444.52 & 2.6   & 1.5   & -8.9  &  -9.0  \\
M31-004522.58 & \nodata  & 1.0    & -6.9   & -7.3   \\ 
Var A\tablenotemark{a}   &  \nodata &  \nodata &  \nodata &  -7.9 \\
N093351  &  \nodata  & 0.45   & -8.8  & -8.8  \\  
B324  & 0.8   & 0.9  & -10.2  & -10.1     \\
N125093  & 1.6  & 0.9   & -8.8   &  -8.9 \\ 
\enddata
\tablenotetext{a}{Var A is optically obscured by about 4 magnitudes of circumstellar dust. See text
for a discussion of its luminosity.} 
\end{deluxetable}

%% file: RMHpreprint.bbl
\begin{thebibliography}{}
\bibitem[Bonanos et al.(2006)]{Bonanos}Bonanos, A. Z. et al. 2006, \apj, 652, 313 
\bibitem[Clark et al.(2012)]{Clark12}Clark, J. S., Castro, N., Garcia, M., et al. 2012, \aap, 541, A146
\bibitem[Cutri et al(2003)]{Cutri}Cutri, R. M., Skrutskie, M. F., Van Dyk, S. et al. 2003, The IRSA 2MASS All-Sky Point Source Catalog, NASA/IPAC Infrared Science Archive 
\bibitem[Davidson et al.(1995)]{KD95}Davidson, K., Ebbets, D., Weigelt, G., Humphreys, R M., Hajian, A. R., Walborn, N. R., \&  Rosa, M. 1995, \aj, 109, 1784  
\bibitem[Davidson(1987)]{KD87}Davidson, K. 1987, \apj, 317, 760
\bibitem[de Jager(1998)]{deJ98}de Jager, C. 1998, \aapr, 8, 145
\bibitem[Dessart et al.(2008)]{Dess08}Dessart, L., Hillier, D. J., Gezari, S., Ba
sa, S. \& Matheson, T. 2009, \mnras, 394, 21 
\bibitem[Drout et al.(2009)]{Drout}Drout, M. R., Massey, P., Meynet, G., Tokarz, S., \& Caldwell, N. 2009, \apj, 703, 441  
\bibitem[Drout et al.(2012)]{Drout2012}Drout, M. R., Massey, P. \& Meynet, G. 
2012, \apj, 750, 97
\bibitem[Fabricant et al(1998)]{Fab98}Fabricant, D. G., Hertz, E. N., Szentgyorgyi, A. H., et al. 1998, Proc. SPIE, 3355, 285
\bibitem[Gamen et al.(2012)]{Gamen}Gamen, R., Walborn, N., Morrell, N., Barba, R. 
\&  Fernandez-Lajus, E.  2012, CBET 3192  
\bibitem[Gorlova et al.(2006)]{Gorlova}Gorlova, N., Lobel, A., Burgasser, A. J., Rieke, G. H., Ilyin, I.,  \& Stauffer, J. R. 2006, \apj, 651, 1130 
\bibitem[Hillier et al.(2001)]{Hillier01}Hillier, D. J., Davidson, K., Ishibashi, K., \& Gull, T. R. 2001, \apj, 553, 837  
\bibitem[Hubble and Sandage(1953)]{HS}Hubble, E. \& Sandage, A. 1953, \apj, 118, 353
\bibitem[Humphreys(1980)]{RMH80}Humphreys, R. M. 1980, \apj, 241, 587 
\bibitem[Humphreys \& Sandage(1980)]{HS80}Humphreys, R. M \& Sandage, A. 1980, \apjs, 44, 319
\bibitem[Humphreys(1983)]{RMH83}Humphreys, R. M. 1983, \apj, 269, 335
\bibitem[Humphreys, Jones, \& Gehrz(1987)]{RMH87}Humphreys, R. M., Jones, T. J. \& Gehrz, R. D. 1987, \aj, 94, 315 
\bibitem[Humphreys, Massey, \& Freedman(1990)]{HMF90}Humphreys, R. M., Massey, P. \& Freedman, W. L. 1990, \aj, 99, 84 
\bibitem[Humphreys \& Davidson(1994)]{HD94}Humphreys. R. M. and Davidson, K. 1994, \pasp, 106, 1025 
\bibitem[Humphreys et al.(1997)]{RMH97}Humphreys. R. M., Smith, N., Davidson, K., et al. 1997, \aj, 114, 2778 
\bibitem[Humphreys, Davidson, \& Smith(2002)]{RMH02}Humphreys. R. M., Davidson, K, \& Smith, N. 2002, \apj, 124, 1026
\bibitem[Humphreys et al.(2006)]{RMH06}Humphreys, R. M., Jones, T. J., Polomski, E., et al. 2006, \apj, 131, 2105   
\bibitem[Humphreys et al.(2011)]{RMH11}Humphreys, R. M., et al. 2011, \apj, 743, 118 
\bibitem[Humphreys et al.(2012)]{RMH12}Humphreys, R. M., Davidson, K., Jones, T. J., Pogge, R. W., Grammer, S. H., Prieto, J. L., \&  Pritchard, T. A. 2012, \apj, 760, 93 
\bibitem[Jones et al.(1993)]{Jones}Jones, T. J., Humphreys, R. M., Gehrz, R. D. et al. 1993, \apj, 411, 323  
\bibitem[Lobel, et al.(2003)]{Lobel}Lobel, A., et al. 2003, \apj, 583, 923 
\bibitem[Massey et al.(1996)]{Massey96}Massey, P., Bianchi, L., Hutchings, J. B., \& Stecher, T. P. 1996, \apj, 469, 629 
\bibitem[Massey and Johnson(1998)]{Massey98}Massey, P. \& Johnson, O. 1998 \apj, 505, 793
\bibitem[Massey et al.(2006)]{Massey06}Massey, P., Olsen, K. A. G., Hodge, P. W. et al. 2006, \aj, 131, 2478   
\bibitem[Massey et al.(2007)]{Massey07}Massey, P., McNeill, R. T., Olsen, K. A. G., et al. 2007, \aj, 134, 2474 
\bibitem[McQuinn et al.(2007)]{McQ}McQuinn, K. B. W., Woodward, C. E., Willner, S. P. et al. 2007, \apj, 664, 850 
\bibitem[Mehner et al.(2013)]{Mehner}Mehner, A., Baade, D, Rivinius, T., Lennon, D. J., Martayan, C., Stahl, O. \& Steffi, S. 2013, \aap, submitted  
\bibitem[Melendez, Bautista \& Badnell(2007)]{Melendez}Melendez, M., Bautista, M. A., \& Badnell, N. R. 2007, \aap, 469, 1203  
\bibitem[Monteverde et al.(1996)]{Mont}Monteverde, M. L., Herrero, A., Lennon, D. J. et al. 1996, \aap, 312, 24
\bibitem[Mould et al.(2008)]{Mould}Mould, J., Barmby, P., Gordon, K., et al. 2008, \apj, 687, 230  
\bibitem[Nieuwenhuijzen et al.(2012)]{Hans}Nieuwenhuijzen, H., De Jager, C., Kolka, I., Israelian, G., Lobel, A., Zsoldos, E., Maeder, A. \&  Meynet, G. 2012, \aap, 546, 105   
\bibitem[Osterbrock and Ferland(2006)]{OF}Osterbrock, D. F. \& Ferland, G. F. 20
06, {\it Astrophysics of Gaseous Nebulae and Active Galactic Nuclei} {University
 Science Books} 
\bibitem[Oudmaijer(1998)]{Rene98}Oudmaijer, R. D. 1998, \aap., 129, 541
\bibitem[Oudmaijer et al.(1996)]{Rene96}Oudmaijer, R. D., Groenewegen, M.A.T., 
 Matthews, H.E., Blommaert, J.A.D., \& Sahu,  1996, \mnras., 280, 1062
\bibitem[Riess et al.(2012)]{M31Ceph}Riess, A. G., Fliri, J., \& Valls-Gabaud, D. 2012, \apj, 745, 156 
\bibitem[Scowcroft et al.(2009)]{M33Ceph}Scowcroft, V., Bersier, D., Mould, J. R., \& Wood, P. R. 2009, \mnras, 396, 1287 
\bibitem[Smartt(2009)]{Smart}Smartt, S. 2009, in Ann. Rev. Astron. \& Astrophys., 47, 63  
\bibitem[Suh(1999)]{Suh}Suh, K.-W. 1999, \mnras, 304, 389 
\bibitem[Thompson et al.(2009)]{Thom}Thompson, T.A., Prieto, J. L., Stanek, K. Z. et al. 2009, \apj, 705, 1364 
\bibitem[Valeev et al.(2009)]{Valeev09}Valeev, A. F., Sholukhova, O. N., \& Fabrika, S. N. 2009, \mnras, 396, L21 
\bibitem[Valeev et al.(2010a)]{Valeev10a}Valeev, A. F., Sholukhova, O. N., \& Fabrika, S. N. 2010a, Astrophysical Bull., 65, 149 
\bibitem[Valeev et al.(2010b)]{Valeev10b}Valeev, A. F., Sholukhova, O. N., \& Fabrika, S. N. 2010b, Astrophysical Bull., 65, 381 
\bibitem[van Genderen et al.(1997a)]{vanG97a}van Genderen, A. M., Sterken, C., \& de Groot, M. 1997, \aap, 318, 81 
\bibitem[van Genderen et al.(1997b)]{vanG97b}van Genderen, A. M., de Groot, M., \& Sterken, C. 1997,
 \aaps, 124, 517  
\bibitem[van Genderen \& Sterken(2002)]{vanG2002}van Genderen, A. M. \& Sterken, C. 2002, \aap, 386,  926  
\bibitem[Wright et al.(2010)]{Wright}Wright, E. L. Eisenhardt, P.R. M., Mainzer, A. K., et al. 2010, \aj, 140, 1868 
\end{thebibliography}
